# Imaging thermal conductivity with nanoscale resolution using a scanning spin probe

Abdelghani Laraoui[1], Halley Aycock-Rizzo[1,2], Yang Gao[1,3,4], Xi Lu[4], Elisa Riedo[1,3,4,*], and Carlos A. Meriles[1,2,*]

**The ability to probe nanoscale heat flow in a material is often limited by lack of spatial resolution. Here, we use a diamond-nanocrystal-hosted nitrogen-vacancy centre attached to the apex of a silicon thermal tip as a local temperature sensor. We apply an electrical current to heat up the tip and rely on the NV to monitor the thermal changes the tip experiences as it is brought into contact with surfaces of varying thermal conductivity. With the aid of a combined AFM/confocal setup, we image phantom microstructures with nanoscale resolution, and attain excellent agreement between the thermal conductivity and topographic maps. The small mass and high thermal conductivity of the diamond host make the time response of our technique short, which we demonstrate by monitoring the tip temperature upon application of a heat pulse. Our approach promises multiple applications, from the investigation of phonon dynamics in nanostructures to the characterization of heterogeneous phase transitions and chemical reactions in various solid-state systems.**

## Introduction

Over the last decade nanoscale thermal transport has been at the centre of widespread interest, partly driven by a number of technologically important applications. For example, as the characteristic dimensions of current electronic devices approach a few tens of nanometres[1], fundamental differences between macro- and nano-scale thermal flow make device operation unreliable and complicate increased circuit integration. Nanoscale thermal transport also plays a critical role in the design of better thermal insulation[2] and enhanced thermoelectric energy recovery[3] as well as in thermal nanolithography[4] and the implementation of advanced magnetic storage devices that use thermal energy to manipulate local magnetic states[5]. Although much progress has been made in the recent past, many fundamental aspects of energy flow in nano-structures are not well understood, partly due to the experimental difficulties associated with high-resolution (< 30 nm) thermal sensing. Moreover, observations are often indirect and rely on models difficult to verify unambiguously.

Here we articulate atomic force and confocal microscopy to demonstrate an alternate route to thermal conductivity imaging with nanoscale resolution. Our approach leverages on the unique properties of the nitrogen-vacancy (NV) centre in diamond, a spin-1 defect that can be individually initialized and readout with the aid of optical and microwave pulses[6]. Thermal changes in the NV vicinity can be detected by monitoring the NV spin resonance frequency, which strongly depends on the system temperature[7-9]. In our experiments we use a diamond-nanocrystal-hosted NV as a nanoscale probe attached to a sharp silicon tip. The latter is part of an atomic force microscope (AFM) cantilever with an integrated heater, which can be used to warm the tip to a predefined temperature above ambient. We obtain an image of the sample thermal conductivity as we record the NV temperature drop upon contact with the substrate for different tip positions in a broad range of model samples. Further, we show that the time response of our spin probe is fast, a consequence of the excellent thermal conductivity and minuscule mass of the diamond host. Since the bulk of the sample substrate remains at a uniform, predefined temperature (ambient conditions in the present case), our technique extends the complementary notion of NV-assisted high-resolution thermometry (i.e., imaging thermal gradients across a substrate[10]) as well as the use of the NV as a nanoscale magnetometer demonstrated recently[11-13].

## Results

**NV-assisted thermal sensing.** A more detailed description of our working geometry is presented in Fig. 1: Our system articulates an AFM and a confocal microscope in a two-sided geometry. We use a thermal cantilever formed by two parallel arms of $n$-doped, electrically conductive silicon connected at one end. The small section of the cantilever above the tip is left purposely undoped and serves as a local heater upon injection of an electrical current. We mount the colour centre onto the tip apex by firmly scanning the AFM probe on a sample of disperse diamond nanocrystals loosely bound to a glass substrate. Van der Waals forces keep the nanoparticle securely attached to the tip surface[14]; images of a sample diamond crystal before and after AFM grafting are shown in Figs. 1b through 1e. We use a high-numerical-aperture objective facing the AFM tip to excite and collect fluorescence from the nanoparticle. Figs. 1f through 1h show successive confocal images recording the NV centre position as we displace the tip over a (semi) transparent substrate.

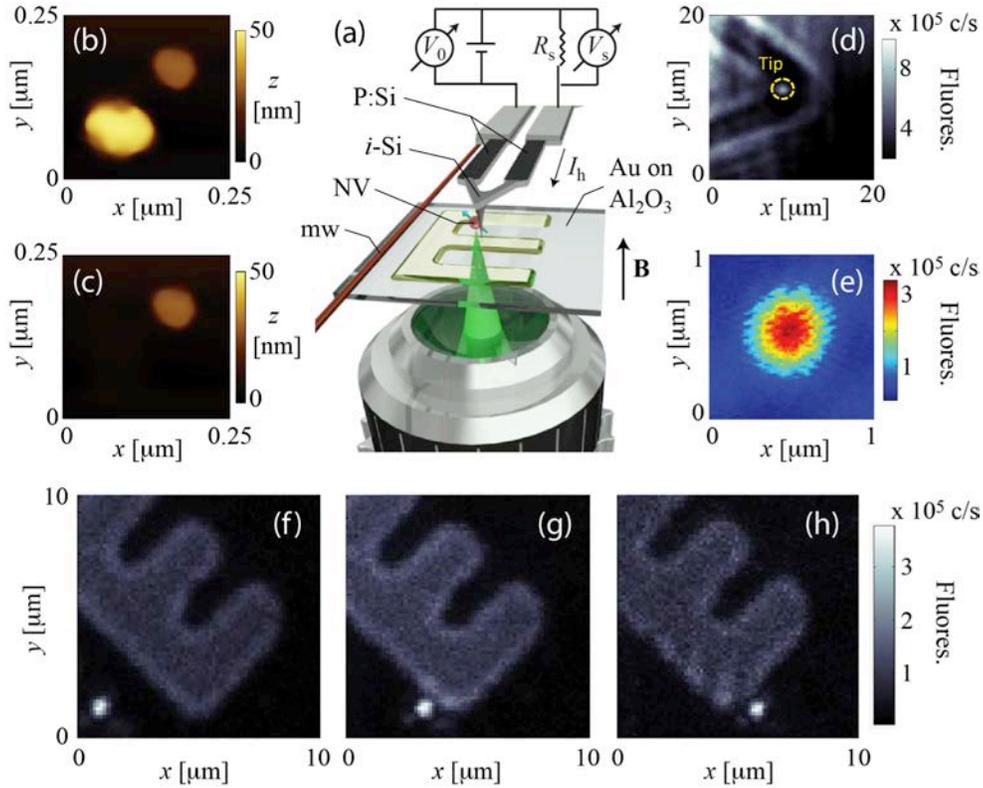

**Fig. 1 | Approach to NV-enabled thermal scanning microscopy.** (a) Schematics of our experimental setup. An electrical current circulates along the arms of a thermal AFM cantilever (phosphorous-doped Si) and heats up the end section above the tip (intrinsic Si). A high-NA objective excites and collects the fluorescence emitted by a diamond-nanocrystal-hosted NV attached to the AFM tip. A wire on the sample surface serves as the source of mw. The resistance (and thus temperature) of the intrinsic segment of the cantilever can be determined from the measured current $I_h = V_s/R_s$ and applied voltage $V_0$. Our experiments are carried out in the presence of a magnetic field **B** along the direction normal to the sample. (b) AFM image of two diamond nanoparticles on a glass substrate. (c) Same as before but after firmly scanning the tip on the larger nanocrystal. Comparison with (b) shows the particle has been removed from the substrate. (d) Fluorescence image of the cantilever end and tip. Light from the tip-attached nanocrystal can be clearly separated from the background (yellow circle). (e) Zoomed confocal image of the NV centre in (d). (f-h) Sequential fluorescence images of the nanoparticle-hosted NV as the AFM scans a micro-structured surface. The mild fluorescence in the back—here serving as a reference—originates from a 18 nm thick, 'E'-shaped pattern of (semitransparent) gold on a sapphire substrate.

In our experiments we apply a voltage difference between the arms of the cantilever to set the heater temperature $T_h$ (see below Methods), which we then determine by monitoring the NV response to microwave. Fig. 2a shows the optically-detected electron spin resonance (ESR) spectra at three different applied voltages, each exhibiting the characteristic NV fluorescence dips near 2.8 GHz[15]. The nanoparticle temperature $T_t$ follows from the known thermal dependence of the NV resonance frequency[7,16] (Fig. 2b). As shown in the figure insert, $T_t$ agrees well with the heater temperature $T_h$ as determined from the circulating current and voltage drop at the cantilever. The latter is to be expected given the relatively homogeneous temperature distribution throughout the intrinsic section of the cantilever[17] (at least in the absence of tip contact with a substrate, see below). It is worth mentioning that the absolute NV fluorescence is also a function of temperature[8,16], which can be exploited as an alternate (arguably simpler) modality of thermal sensing. This route, however, must be excluded in our present experiments given the two-sided geometry of our setup and concomitant dependence of the light collection efficiency on the tip position. Further, we show (Supplementary Note 1) that spin-assisted thermal sensing can attain significantly higher detection sensitivity, particularly given the ability to engineer nanoparticle-hosted NVs with near-bulk spin lifetimes[18].

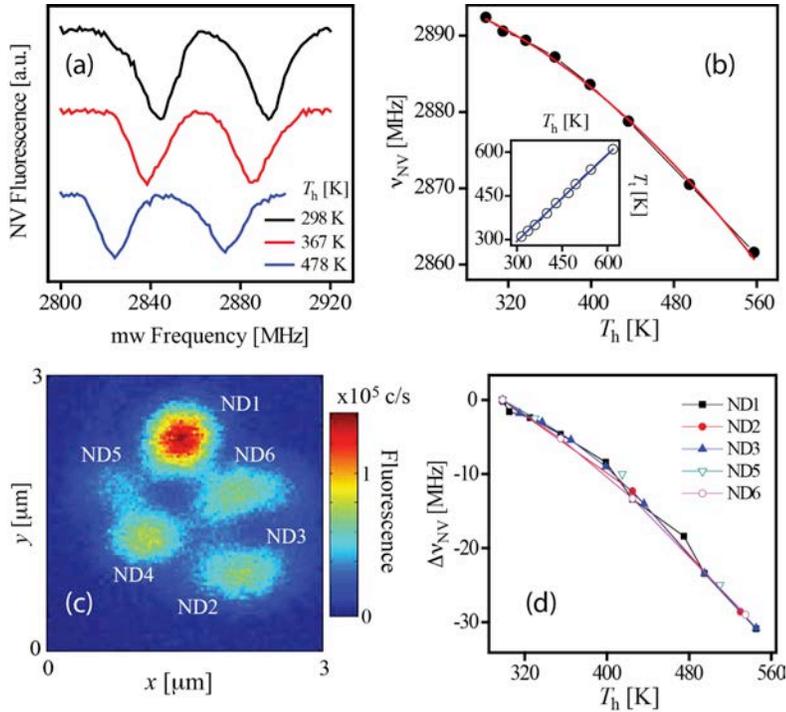

**Fig. 2 | NV-spin-assisted sensing of the AFM tip temperature.** (a) Optically-detected magnetic resonance spectra of an NV centre attached to the thermal tip at three different heater temperatures $T_h$. The right (left) dip corresponds to the $m_S = 0 \leftrightarrow m_S = +1$ transition (the $m_S = 0 \leftrightarrow m_S = -1$ transition) between the levels of the NV ground-state spin triplet. The line splitting originates from a ~1 mT magnetic field aligned along the NV axis. Spectra have been displaced vertically for clarity. (b) Temperature dependence of the $m_S = 0 \leftrightarrow m_S = +1$ transition frequency $\nu_{NV}$; the red trace is a quadratic fit. $T_h$ is determined from the resistance of the intrinsic section of the cantilever (see Methods). The insert shows the diamond nanoparticle temperature $T_t$ using the known relation with the NV resonance frequency as a function of the heater temperature $T_h$ for a tip exposed to ambient. Comparison with the graph diagonal (solid blue trace) indicates that both coincide within experimental error. (c) Confocal image of the AFM cantilever in the vicinity of the tip after grafting multiple NV-hosting nanocrystals. (d) Frequency shift $\Delta\nu_{NV}$ relative to room temperature for each of the nanocrystals imaged in (c).

Interestingly, we find that more than one diamond particle can be attached to the AFM probe if, once loaded, the tip is brought into contact with another NV-hosting diamond crystal. This is shown in Fig. 2c where we attach up to six nanocrystals upon repeated grafting. We find that the diamond particles decorate the tip apex more or less concentrically (Fig. 2c). Grafting a new particle (ND1, bright spot in the figure) pushes the rest farther deep and into the cantilever body allowing us to optically separate most NVs. Given the random orientation of each particle, we discriminate between NVs close to each other via the application of a small magnetic field. For a given applied voltage, we find identical frequency shifts in all NVs (Fig. 2d), indicative of a consistent thermal contact between the tip and the spin probe.

**Characterization of sample substrates.** Thermal energy is extracted from the tip when physical contact is made with a substrate at a lower temperature. Ignoring from now surface adsorbates, the transport rate depends on the thermal conductance of the substrate: Conductive materials efficiently redistribute the excess energy locally injected by the tip thus leading to a comparatively high heat transport rate. The opposite regime applies to insulating systems, whose temperature increases quickly near the tip point of contact, hence reducing any subsequent thermal flow. For a constant electrical power dissipated at the cantilever end, the immediate consequence is a sample-dependent temperature drop at the tip that can be monitored locally with the aid of the probe NV. This is shown in Fig. 3a where we measure the NV frequency shift (relative to room temperature) as a function of the applied voltage for substrates of varying thermal conductivity. In all cases we find a quadratic response but the scaling depends on the sample material. In particular, the frequency shift is minimum when the tip is in contact with bulk diamond, the system with the highest thermal conductivity.

Figure 3b shows the same data set, this time expressed as the tip temperature $T_t$ and plotted as a function of the material thermal conductivity. The near-logarithmic dependence unveils an increasingly lower contrast between materials whose thermal conductivities are comparatively high. We interpret this observation as a direct consequence of the finite thermal conductivity of intrinsic silicon, which sets an upper limit on the rate of energy transfer from the main heater body into the tip. As the substrate thermal conductivity $\kappa_s$ increases, the (steady-state) tip temperature asymptotically approaches a limit value thus making the system incrementally less sensitive to changes in the substrate ability to transport heat. Implicit in this picture is the appearance of a temperature gradient between the heater body and the AFM tip, a notion we confirm by comparing the temperatures inferred from the voltage drop or the NV frequency shift (Fig. 3c). Unlike Fig. 2, we find that $T_t$ consistently takes a value lower than the heater temperature $T_h$, the difference growing with the substrate conductivity. For example, for $T_h = 500$ K, contact with a diamond crystal brings the tip temperature down by about 100 K. This behaviour can be captured quantitatively via a model that takes into account the thermal resistance of the tip, the sample, and the interface (see Supplementary Note 2 and Refs. [19] and [20]).

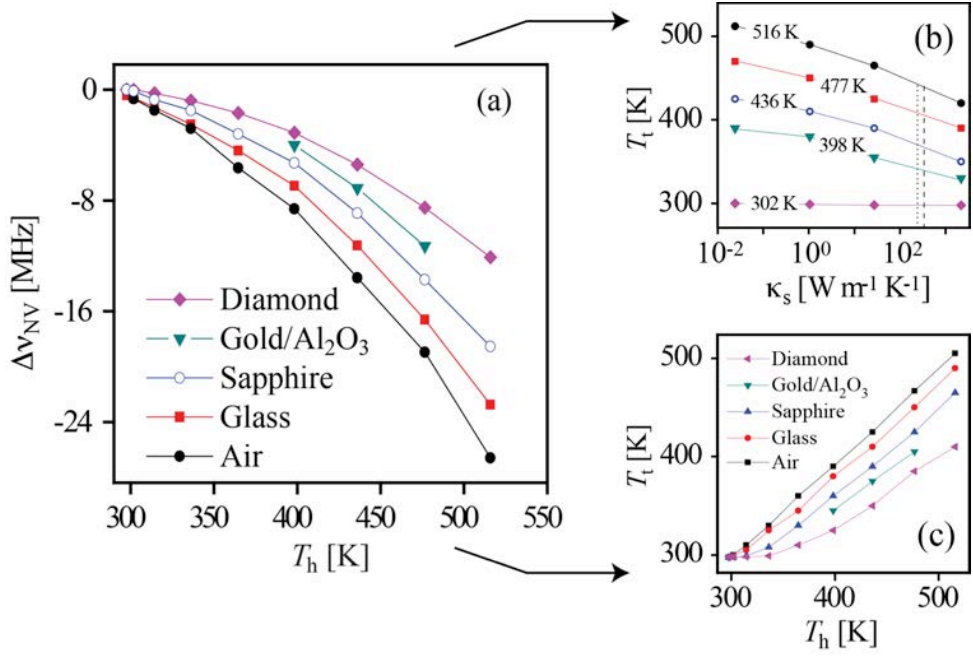

**Fig. 3 | Influence of the substrate on the tip temperature.** (a) NV frequency shift $\Delta\nu_{NV}$ relative to ambient as a function of the heater temperature $T_h$ when exposed to air or upon contact with substrates of different composition. The system-dependent NV shifts at a fixed heater temperature indicate contact-induced selective cooling of the AFM tip. (b) Tip temperature $T_t$ versus the thermal conductivity $\kappa_s$ for various heater temperatures $T_h$ when the heater is exposed to air ($\kappa_s \sim 2\times 10^{-2}$ W m$^{-1}$ K$^{-1}$), or in contact with glass ($\kappa_s \sim 1$ W m$^{-1}$ K$^{-1}$), sapphire ($\kappa_s \sim 35$ W m$^{-1}$ K$^{-1}$), or diamond ($\kappa_s \sim 2\times 10^3$ W m$^{-1}$ K$^{-1}$). Within the range explored, we find $\Delta\nu_{NV} \sim A_{\kappa_s}(T_h)\ln(\kappa_s/\kappa_t(T_h))$ where $A_{\kappa_s}(T_h)$, $\kappa_t(T_h)$ are functions of the heater temperature $T_h$ (and tip thermal conductivity, see Supplementary Note 1). Using the data in (a), we estimate the thermal conductivity of 18 nm gold on sapphire to be 230±20 Wm$^{-1}$K$^{-1}$ (vertical dotted line), smaller than 314 Wm$^{-1}$K$^{-1}$, the accepted value for bulk gold (vertical dashed line). (c) Tip temperature $T_t$ versus the heater temperature $T_h$ for different substrates. In (a) and (c) 'Gold/Al$_2$O$_3$' indicates a 18-nm-thick gold film on sapphire.

Building on the sensitivity of the nanodiamond-hosted NV we record the tip temperature as we scan a phantom structure formed by a patterned, 18-nm-thick film of gold on sapphire, shown for reference in Fig. 4a. This image effectively corresponds to a heat dissipation map, largely correlated with the substrate thermal conductivity if the thermal impedance at the point of contact is substrate-insensitive (Supplementary Note 2). Compared to the AFM image, the coarser pixel size (~50 nm) is a consequence of our presently asynchronous operation of the atomic force and confocal microscopes (the tip must be independently positioned prior to NV readout) and hence does not reflect on the technique's spatial resolution, ultimately defined by the AFM tip radius (10 nm). This ideal limit, however, can only be reached as the heater-substrate temperature difference $\Delta T_{hs} \equiv |T_h - T_s|$ — and, correspondingly, the effective sample volume — is brought to a minimum. The latter is demonstrated in Fig. 4c displaying one-dimensional cross sections of the sample gold structure at different heater temperatures; the measured thermal conductivity follows the sample composition more closely for the lower value of $\Delta T_{hs}$, though at the expense of a partial sensitivity loss (see below). This behaviour exposes a trait common to most forms of scanning microscopy, namely, spatial resolution and detection sensitivity are inversely related.

Of note, the image in Fig. 4b shows some patches of low (high) thermal conductance within (outside) the contour of the gold structure that do not correlate with the local topography (compare, e.g., circled areas in Figs 4a and 4b). We hypothesize that the discrepancy may stem from a thin (<2 nm) residual layer of photoresist (gold) after completion of the pattern (see below Methods). This notion is consistent with the critical role assigned to the interface thermal impedance in the accepted models of energy flow between AFM tips and substrates in contact[21], and, if confirmed, could be exploited to highlight slight changes in the surface composition. Interestingly, regions of the gold structure seemingly devoid from surface contamination show at best $\kappa_s^{(Au\ film)} = 230 \pm 20$ Wm$^{-1}$K$^{-1}$, significantly below 314 Wm$^{-1}$K$^{-1}$, the accepted value for bulk gold[22] (see vertical lines in Fig. 3b). This observation is in qualitative agreement with prior thermoreflectance studies (reporting $\kappa_s^{(Au\ film)} = 155 \pm 20$ Wm$^{-1}$K$^{-1}$ for 20 nm gold films on fused silica[23]), and suggests the possibility of probing size effects on phonon transport with the spatial resolution inherent to tip-based scanning techniques.

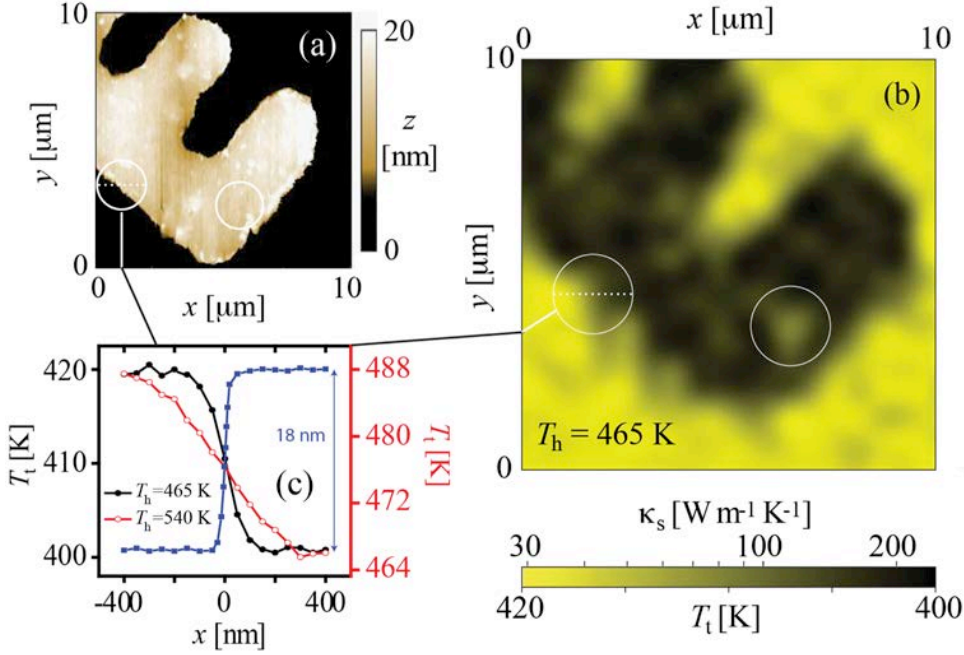

**Fig. 4 | High-resolution thermal conductivity imaging.** (a) AFM topographic image of a 18 nm thick gold structure on sapphire. (b) NV-assisted thermal conductivity image of the same structure. The heater temperature $T_h$ is 465 K. White circles in (a) and (b) indicate example patches of low (high) conductivity inside (outside) the gold structure that do not correlate with the substrate topography. (c) Measured tip temperature near the edge of the gold structure at two different heater temperatures, 465 K (full black circles, left vertical axis) and 540 K (empty red circles, right vertical axis). The topographic curve displaying the 18-nm-thick gold film edge as measured with the AFM (blue squares) is also present for reference. As expected, the spatial resolution of the thermal measurements worsens at higher heater temperatures.

The metric that quantitatively gauges the system sensitivity to the substrate thermal conductivity is given by $\eta_{\kappa_s} \equiv (\Delta \kappa_s)_{min} \sqrt{t_{tot}}$, where $(\Delta \kappa_s)_{min}$ represents the minimum detectable change of thermal conductivity and $t_{tot}$ is the total experimental time. Assuming shot-noise-limited detection, we find (see Supplementary Note 1 and Refs [24], and [25])

$$\eta_{\kappa_s} \sim \frac{2}{\sqrt{\alpha(T_t)\, t_{2NV}}} \left| \left(\frac{d\nu_{NV}}{dT_t}\right)\left(\frac{dT_t}{d\kappa_s}\right) \right|^{-1}, \quad (1)$$

where $\alpha(T_t)$ is the average number of photons collected during a given measurement time when the NV spin is in the $m_S = 0$ state, and $t_{2NV}$ is the NV spin transverse relaxation time. As expected, improved sensitivity is attained for enhanced photon counts and longer-lived NVs. The latter is possible through the use of diamond crystals of higher purity[18], or the implementation of multi-pulse thermal sensing sequences[8,9] (see also Supplementary Fig. 4). To enhance $\alpha$ one can resort to improved photon collection tactics or to nano-crystals hosting multiple NVs (as demonstrated recently for single cell thermal sensing[26]).

A detailed analysis of Eq. (1) allows us to identify the optimum heater temperature $T_h^{(opt)}$, which emerges from a complex interplay between the rate of frequency change, the NV fluorescence, and the substrate thermal conductivity. For our present conditions we calculate $T_h^{(opt)} \sim 620 - 660$ K, depending on the exact value of $\kappa_s$, which corresponds to a tip temperature hovering around $T_t^{(opt)} \sim 570$ K. Near this optimum we find $\eta_{\kappa_s}^{(opt)}/\kappa_s \sim 5$ % Hz$^{-1/2}$ for the lifetime $t_{2NV} = 500$ ns characteristic of the NVs studied herein; relative sensitivities of order 0.2 % Hz$^{-1/2}$ are anticipated, however, given the longer NV lifetimes (approaching 200 µs) possible in higher-quality nanocrystals via the use of dynamical decoupling protocols[18]. Notice that despite the local sensitivity minimum at $T_h^{(opt)}$, the practical temperature range of operation is considerably broad (Supplementary Note 1). For example, at a heater temperature of 400 K — ~250 K below the optimum and only ~100 K above ambient conditions — the sensitivity loss relative to $\eta_{\kappa_s}^{(opt)}$ amounts to approximately a factor 2.

In the ideal limit where the thermal impedance at the point of contact is insensitive to the substrate thermal conductivity — Supplementary Note 2 — our ability to derive a well-defined relation between $T_t$ and $\kappa_s$ rests on the implicit assumption of a known, uniform substrate temperature. Strictly, however, this latter condition is never met because the finite thermal conductivity of the substrate produces a thermal gradient between sections of the sample close to or removed from the tip, thus introducing an uncertainty in the substrate temperature. The concomitant error in the value of $\kappa_s$ is relatively small in the systems studied herein — where the thermal conductivity does not show a strong temperature dependence — but could play a non-negligible role in some specific cases, e.g., when the tip is hot enough to locally induce a phase transition. Naturally, this problem can be mitigated by reducing the temperature difference between the heater and bulk substrate, though at the expense of a partial sensitivity loss as discussed above. In the opposite regime, enhancing the heater-induced thermal gradient can be exploited as a tool to introduce alternate forms of contrast, particularly in situations where, rather than the exact value of $\kappa_s$, one is interested in exposing, e.g., compositional sample heterogeneity.

Another aspect of practical interest concerns the influence of the heater current on the NV resonance frequency, and thus on our ability to correctly determine the temperature drop at the tip. A quick inspection of the heater characteristics (see Methods and Supplementary Fig. 2) shows that the current required to bring the tip to 500 K amounts to about 1 mA. A crude estimate for the present heater geometry (in the form of ~1 µm thick, 20 µm wide V-shape, see Fig. 1a) yields a magnetic field $B_c$ of order 10 µT along an axis perpendicular to the tip, which translates in a frequency shift of up to ~300 kHz for a properly oriented NV. Though significantly smaller than the NV line width herein, comparable shifts can be easily picked up, particularly if longer-lived NVs are used as the probe. We note, however, that because the tip makes a marginal contribution to the total heater mass, the cantilever current is insensitive to a temperature drop in the tip. Correspondingly, $B_c$ has no impact on the recorded images other than a minor readout shift uniform throughout the scan. Further, this shift is negligible if the NV axis and the field direction are perpendicular, as in the present case.

**Time response.** The small mass and high thermal conductivity of the diamond nanocrystal may be exploited to probe fast, time-dependent physical processes impacting the energy flow between the tip and the substrate (e.g., transient heat waves or fast phase transitions[27]). As a proof of principle, we apply a square voltage pulse (rise/decay time of 4 ns) between the two conducting arms of the cantilever, and use the NV centre to monitor the corresponding temperature change at the tip. The pulse sequence is presented in Fig. 5a: To read $T_t$ at a given time, we collect the NV fluorescence upon NV initialization via a 1-µs-long pulse of green light and a 500-ns inversion pulse of microwave (mw) resonant with the NV $m_S = 0 \leftrightarrow m_S = 1$ transition at 436 K. Given the spectral selectivity of the mw pulse, the spin inversion is conditioned on the diamond nanocrystal temperature thus making the NV fluorescence sensitive to the applied voltage.

The result is presented in Fig. 5b for the case of a 6-ms-long voltage pulse; the NV fluorescence is initially bright because the mw is off resonance at room temperature and hence has no effect on the NV spin. Upon current injection at $t = 6$ ms, the cantilever reaches the steady-state temperature $T_h = 436$ K, which is accompanied by a concomitant reduction of the NV fluorescence upon spin inversion. The transition between both stationary states is exponential and has a characteristic response time $t_d = 184$ µs (insert to Fig. 5b). This value is in good agreement with prior studies on similar thermal AFM probes[28], and is determined by the heat capacity of the intrinsic section of the cantilever and the probe reactance. We emphasize that $t_d$ must not be confused with the time resolution, ultimately defined by the NV spin readout time $t_r \sim 1$ µs. The latter is much longer than the time $t_l$ it takes the tip (and the adjacent diamond nanocrystal) to thermalize with the substrate. For example, assuming a distance $\lambda \sim 100$ nm between the 50-nm nanocrystal and the tip apex, a crude estimate for a silicon tip yields $t_l \sim \lambda^2 \rho_t c_t / \kappa_t \sim 100$ ps, where we have used the room temperature values of the tip thermal conductivity $\kappa_t$, specific heat $c_t$, and density $\rho_t$. Note that since the thermal conductivity of diamond is about ten times greater than silicon, the nanoparticle shares the tip temperature at all times. Such a short time scale could find use, e.g., in the characterization of the thermal response of complementary metal-oxide semiconductor (CMOS) devices, of order 10 ns[29]. Given that $t_l \ll t_r$, however, higher photon collection efficiency and adapted sensing protocols will be mandatory to access this ultrafast regime.

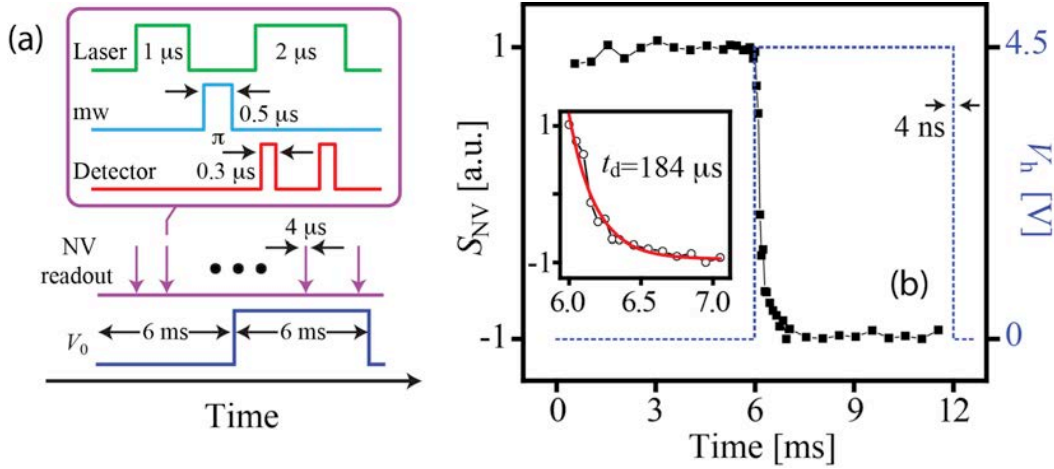

**Fig. 5 | Probing time-dependent thermal processes.** (a) Pulse sequence. We induce a temperature transient in the tip by applying a sharp (4 ns rise/decay time) voltage pulse across the thermal cantilever. We monitor the tip temperature by reading the NV spin state at different points during the application of the voltage pulse (purple arrows). The readout protocol (upper square) relies on the fluorescence collected upon NV initialization (1 μs laser pulse) and application of a selective π-pulse (0.5 μs of mw) resonant with the NV $m_S = 0 \leftrightarrow m_S = 1$ transition at $T_h = 436$ K (corresponding to $V_0 = 4.5$ V). The photo-detector is opened twice during the readout, the first time to collect the spin-dependent fluorescence (300 ns read) and the second one to acquire a reference (300 ns) after the NV re-pumps into $m_S = 0$. (b) NV response $S_{NV}$ (left vertical axis) as a function of time; the scale is normalized relative to the maximum NV fluorescence contrast. The signal goes from bright to dim as the voltage pulse (dashed blue line) reaches its maximum (right vertical axis) and the NV tunes to resonance with the mw pulse. The figure insert zooms on the NV signal transition around 6 ms; we find a time constant $t_d = 184$ μs.

**Discussion**

While the experiments herein are conceived to measure the temperature drop in the thermal tip upon contact with a cooler substrate, the converse approach is also conceivable, namely, an NV near the apex of a tip in a room temperature cantilever can be used to reconstruct the temperature map of a hot substrate. This configuration could be exploited, for example, to identify and characterize the 'hot spots' forming at the junctions of semiconductor structures, known to play a central role in the generation of heat within integrated electronic devices. By the same token, pairs of NV-hosting AFM tips — one serving as a heat source, the other as a local temperature sensor — could provide new opportunities for the investigation of heat transport in nanostructured systems with high spatial resolution. Naturally, this notion can be extended to include systems of multiple, arrayed thermal cantilevers already developed for high-density thermo-mechanical data storage[30].

Of note, physical contact between the thermal tip and the substrate is not a mandatory condition, particularly if a tuning fork is used to control the distance to the sample surface with high precision. For example, ~1 nm gaps between the tip and the surface provide an ideal platform for the investigation of radiative transport at the nanoscale[31]. Along these lines, transition to a single-sided geometry — already implemented in recent demonstrations of nanoscale NV magnetometry — can extend the present result to the more general class of opaque materials, precluded in this initial demonstration. Likewise, NV centers engineered within diamond films overgrown on a thermal AFM probe provide an alternate, arguably more robust platform than the grafted nanoparticle presented herein, particularly if precision implantation[32] is used to control the NV position relative to the tip apex. This class of sharp (sub-10 nm) NV-hosting thermal tips could be applied to a range of outstanding problems including the understanding of heat transport across interfaces connecting dissimilar materials, or to the investigation of phonon dynamics in narrow gaps and in molecular or 2D systems weakly coupled to the substrate[33].

## Methods

***Sample preparation:*** To carry out the experiments of Fig. 4 we used UV lithography to create a gold grid (314 Wm$^{-1}$K$^{-1}$) on a 80 μm thick sapphire substrate (30 Wm$^{-1}$K$^{-1}$). Sapphire was spin coated with 1:2 NR9-1000PY:IPA for 3 minutes at 6100 rpm, soft-baked atop a preheated 130 μm thick glass cover slip at 90 ºC for one minute, exposed with a chromium pattern by UV lamp for 5 seconds, hard-baked at 100 ºC on the same cover slip for one minute, and then developed in RD6 for 7 seconds before rinsing with deionized water. Gold was sputtered on to the resist pattern for 180 seconds at 20 mA and 45 mTorr and then ultra-sonicated in acetone for 5 seconds for lift-off. The resulting Au grid lines, separated by 100 μm, had nodes individually labeled by four coordinate numbers. Each numeric symbol has a line width of 2 μm. The pattern is about 18 nm thick and nearly optically transparent, which renders the structure compatible with the two-sided geometry of our confocal/AFM setup. The experiment in Fig. 4 was done on a particular coordinate number with ideal optical characteristics (see Supplementary Fig. 1).

***Operation of the thermal AFM tip:*** We control the heater temperature $T_h$ via the circuit sketched in Fig. 1a. We use a power supply to induce a voltage difference $V_0$ up to 20 V; the series resistance $R_s = 2$ kΩ protects the cantilever from current spikes and allows us to determine the current $I_h = V_s/R_s$ circulating through the heater via the measured voltage drop $V_s$. We determine the heater resistance $R_h$ via the relation

$$R_h(T_h) = R_s \frac{(V_0 - V_s)}{V_s} \qquad (2)$$

where we made explicit the dependence on the heater temperature $T_h$. The latter, in turn, can be controlled via the power $P_h$ dissipated at the cantilever given by

$$P_h = V_s \frac{(V_0 - V_s)}{R_s} \ . \qquad (3)$$

In Supplementary Fig. 2a we plot $R_h$ as a function of the dissipated power $P_h$. Below the critical value $P_h^{(c)} = 8.6$ mW the cantilever resistance increases monotonically from about ~1 kΩ, the value at room temperature, to about 4.5 kΩ. This increase reflects the reduced mobility of the conduction electrons in the doped legs of the cantilever at higher temperatures. The sharp decay at higher applied voltages is a direct consequence of the abrupt thermal activation of silicon carriers at $T_h^{(c)} = 823$ K, which provides an internal temperature reference. On the other hand, in situ Raman spectroscopy shows[34] that the heater temperature can be estimated from the linear relation

$$T_h = T_a + \left(T_h^{(c)} - T_a\right)\frac{P_h}{P_h^{(c)}} \ , \qquad (4)$$

where $T_a$ denotes ambient temperature. Therefore, combining the data from Supplementary Fig. 2a and Eqs. (3) and (4), we can generate a calibration curve between the applied voltage $V_0$ and the cantilever temperature $T_h$ (Supplementary Fig. 2b).

**Acknowledgements**

We thank Prof. W.P. King at the University of Illinois, Urbana-Champaign for kindly providing the thermal tips used in our experiments. A.L., H.A-R., and C.A.M. acknowledge support from the National Science Foundation through grants NSF-1401632 and NSF-1309640. Y.G. and E.R. acknowledge the support of the Office of Basic Energy Sciences of the US Department of Energy (DE-FG02-06ER46293). Y.G. acknowledges the support of the National Science Foundation (NSF) grant CMMI-143675.

**Author contributions**
A.L. performed the measurements on thermal conductivity, with the assistance of X.L.; H.A-R. fabricated the sample microstructures and assisted with some of the imaging measurements. Y.G. and E.R. developed the model for tip-surface heat transport. C.A.M., A.L., and E.R. conceived the experiments and analyzed the data. All authors wrote and commented on the manuscript.